\documentclass{egpubl}
\usepackage{eurovis2023}
\usepackage{booktabs}
\usepackage{tabularx}
\usepackage{xcolor, colortbl}
\usepackage{array}

\SpecialIssuePaper         

\CGFStandardLicense

\usepackage[T1]{fontenc}
\usepackage{dfadobe}  
\usepackage{fontawesome}

\usepackage{cite}  
\BibtexOrBiblatex
\electronicVersion
\PrintedOrElectronic
\ifpdf \usepackage[pdftex]{graphicx} \pdfcompresslevel=9
\else \usepackage[dvips]{graphicx} \fi

\usepackage{egweblnk}

\newcommand{\envicon}[0]{\hspace{1mm} \includegraphics[width=.015 \textwidth]{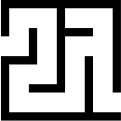}\hspace{0.75mm}}

\newcommand{\hagenticon}[0]{\hspace{0.25mm} \includegraphics[width=.015 \textwidth]{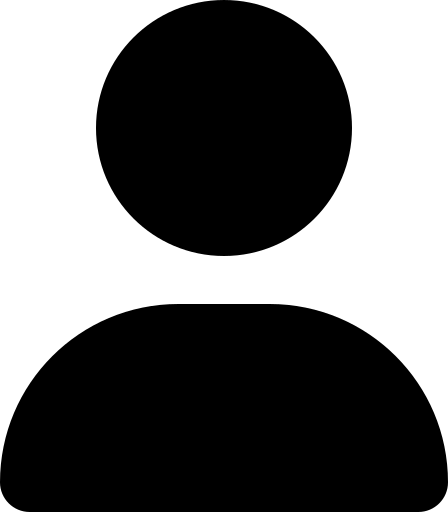}\hspace{0.75mm}}

\newcommand{\cagenticon}[0]{\hspace{0.25mm} \includegraphics[width=.015 \textwidth]{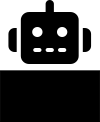}\hspace{0.25mm}}

\newcommand{\acro}[1]{\textsc{\MakeLowercase{#1}}}

\definecolor{alloyorange}{rgb}{0.749,0.341,0}

\title
      {
      Human--Computer Collaboration for Visual Analytics:\ \\ an Agent-based Framework}

\author[S.\ Monadjemi, M.\ Guo, D.\ Gotz, R.\ Garnett, A.\ Ottley]
{\parbox{\textwidth}{\centering 
        Shayan Monadjemi$^{1}$\orcid{0000-0002-9385-5969},
        Mengtian Guo$^{2}$\orcid{0000-0002-9299-7594},
        David Gotz$^{2}$\orcid{0000-0002-6424-7374},
        Roman Garnett$^{1}$\orcid{0000-0002-0152-5453},
        Alvitta Ottley$^{1}$\orcid{0000-0002-9485-276X}
        }
        \\
{\parbox{\textwidth}{\centering 
         $^1$ Washington University in St. Louis, St. Louis, MO \\
         $^2$ University of North Carolina at Chapel Hill, Chapel Hill, NC
       }
}
}

\begin{document}


\maketitle
\begin{abstract}
The visual analytics community has long aimed to understand users better and assist them in their analytic endeavors. 
As a result, numerous conceptual models of visual analytics aim to formalize common workflows, techniques, and goals leveraged by analysts. 
While many of the existing approaches are rich in detail, they each are specific to a particular aspect of the visual analytic process.
Furthermore, with an ever-expanding array of novel artificial intelligence techniques and advances in visual analytic settings, existing conceptual models may not provide enough expressivity to bridge the two fields.
In this work, we propose an agent-based conceptual model for the visual analytic process by drawing parallels from the artificial intelligence literature.
We present three examples from the visual analytics literature as case studies and examine them in detail using our framework.
Our simple yet robust framework unifies the visual analytic pipeline to enable researchers and practitioners to reason about scenarios that are becoming increasingly prominent in the field, namely mixed-initiative, guided, and collaborative analysis.
Furthermore, it will allow us to characterize analysts, visual analytic settings, and guidance from the lenses of human agents, environments, and artificial agents, respectively.


\end{abstract}  

\section{Introduction}

In visual analytics (\acro{VA}), humans and computers apply their respective strengths to solve problems involving data. 
Understanding this human--computer partnership for data analysis is a vast and complex process due to its interdisciplinary nature \cite{keim2008visual}. 
On one hand, we need to better understand the cognitive and perceptual processes involved in how humans analyze data.
On the other hand, we need innovative techniques that transform raw data into actionable knowledge.
Therefore, \acro{VA} researchers have adopted a divide-and-conquer approach in their investigations with the vision that their findings will unite to fully realize this overarching goal of human--computer collaborative analysis. 
Over the years, these investigations have prompted a set of theoretical questions such as \emph{what is the purpose of visual analytics} \cite{andrienko2018viewing}, \emph{how do humans generate knowledge} \cite{sacha2014knowledge}, and \emph{what are the strengths of humans and computers in collaborative data analysis} \cite{crouser2012affordance}. 
Researchers have offered their answers to these questions as a set of \emph{conceptual frameworks} which have evolved over time.
Meanwhile, these frameworks have provided the research community with a way to organize their investigations using a common language.

The existing frameworks have focused on how \emph{human intelligence} (e.g.\ visual perception, social abilities, etc.) can complement \emph{computational powers} (e.g.\ storage, data processing, etc.) for more effective data analysis \cite{crouser2012affordance}.
In recent years, however, advances in the field of artificial intelligence (\acro{AI}) have enabled computers to act as intelligent entities in various settings such as drug and material discovery \cite{mukadum2021efficient}.
At the same time, larger amounts of data and the growing number of analytic techniques have resulted in more complex \acro{VA} scenarios than before. 
These advances in \acro{AI} and rising complexities in \acro{VA} settings have further ignited the desire to design mixed-initiative visual analytic systems which identify the user's analytic intents (e.g., \ \cite{ottley2019follow,monadjemi2020competing}) and take actions to contribute to the process (e.g., \ \cite{kim2019topicsifter, sperrle2021learning, wall2017podium,monadjemi2022guided}).
While existing frameworks of \acro{VA} are rich in detail, we argue that they lack expressivity to reason about the entire mixed-initiative \acro{VA} pipeline.

To address this gap, we need a common language to reason about the visual analytic environment (e.g.\ data, interfaces, etc.) and every potential entity taking actions upon the environment (e.g.\ analysts, automated analytic techniques, etc.).
Therefore, we propose an agent-based conceptual framework by drawing parallels from the \acro{AI} literature. 
In much of \acro{AI} research, problems are reduced to agents interacting with their environment by making observations and taking actions. 
As evident, however, this conceptual framework is abstract and needs to be instantiated for each application area. 
For example, one instantiation of this framework may involve a robot vacuum cleaner that observes its environment consisting of the walls and furniture using its impact sensors, and in return uses its wheels and brushes to turn, move, and clean the floors efficiently.
We aim to view \acro{VA} settings from a similar agent-based perspective.

In this work, we first outline an instantiation of this agent-based framework for visual analytics where human agents (e.g.\ analysts) interact with a visual analytic environment (e.g.\ data, visual interface, etc.) in order to perform their analytic tasks. 
Furthermore, artificial agents (e.g.\ automated processes, guidance engines, etc.) may also interact with the visual analytic environment to take autonomous actions or guide their human counterparts.
Once we define visual analytic agents and environments, we discuss some of their attributes using the same vocabulary as the \acro{AI} community.
We envision that by using a common language, we can bridge \acro{AI} and \acro{VA} research more effectively, making emerging \acro{AI} techniques more accessible to our upcoming investigations of mixed-initiative \acro{VA}.

Second, we consider the existing classification of artificial agents according to their behavior types and apply a similar classification to characterize  higher-level reasoning in analysts. 
An example is the class of \emph{goal-based} agents who take actions in pursuit of a pre-defined goal state.
In contrast to existing \acro{VA} taxonomies which describe \emph{what} analysts do (e.g.\ \cite{gotz2009characterizing, brehmer2013multi}), we believe our approach takes us a step closer towards answering \emph{why} analysts take certain actions.
While we acknowledge that this characterization is a simplification of how humans exhibit higher-level reasoning, we argue that it offers a balanced trade-off between characterizing humans as unknown black boxes (e.g.\ \cite{crouser2016toward}) and treating humans as intractably complex beings to characterize.

Third, we frame visual analytic guidance as \emph{artificial agents} who are able to reason about the analytic process and take actions in tandem with the analysts. 
From this perspective, developing \acro{VA} guidance is the process of designing effective artificial agents, enabling them to reason about the environment and analysts, and empowering them to take appropriate actions at the right time. We believe recent advances in mixed-initiative \acro{VA} systems fit this perspective. For example, user modeling techniques enable artificial agents to reason about analysts by observing their low-level interactions \cite{ha2022unified} and recommendation systems take actions to assist analysts in exploration \cite{sperrle2021learning, monadjemi2022guided}.

To demonstrate how our framework can be applied to visual analytic settings, we present three case studies from the literature.  
Upon reviewing the related work, we selected these examples because they, among other candidates, provide enough details about the system and the analytic task for us to offer an in-depth characterization.
For each case study, we highlight the human agents, the components of the environment, and the artificial agents (when present). Then, we discuss how and if the analyst may exhibit each class of the high-level behaviors while performing the task.

We conclude by offering an in-depth discussion on the areas of focus and gaps for further investigations in mixed-initiative visual analytic settings. Furthermore, we discuss some limitations and recommended extensions of this agent-based framework.


\section{Background}
\label{s:background}

Prior to presenting our conceptual framework, we provide a brief background that spans both fields of visual analytics (\acro{VA}) and artificial intelligence (\acro{AI}). In particular, we focus on the following topics which are most relevant: \emph{agent-based models}, \emph{mixed-initiative visual analytics}, and \emph{conceptual frameworks for visual analytics}.

\subsection{Agent-based Models}

\begin{figure}[!h]
    \centering
    \includegraphics[width=\linewidth]{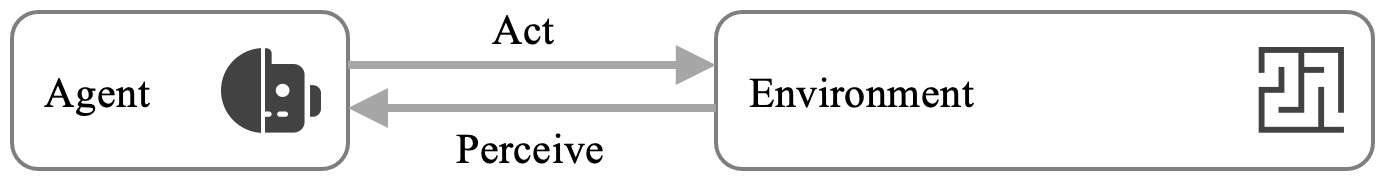}
    \caption{Basic agent-based model from AI literature \cite{russellnorvigaibook}.}
    \label{fig:basic_agent_environment}
\end{figure}

Agent-based models are based on the simple idea that \emph{agents} iteratively interact with their \emph{environment} to serve their purpose. This interaction consists of perceiving the state of the environment through sensors and taking actions via actuators (Fig.\ \ref{fig:basic_agent_environment}).

An \emph{agent} is one who acts or exerts power \cite{mw:agent}. Building on this basic definition, \acro{AI} researchers have defined agents as ``anything that can be viewed as perceiving its environment through sensors and acting upon that environment through actuators'' \cite{russellnorvigaibook}. More specifically, an \emph{intelligent} agent is an agent that exhibits the following types of behavior in order to meet their delegated objectives: proactiveness (taking initiative), reactivity (reacting to changes in a timely manner), and social ability (interact with other agents) \cite{wooldridge1995intelligent, weissmas}.

An \emph{environment} is the collection of circumstances, objects, and conditions by which one is surrounded \cite{mw:environment}. 
Since specific aspects of an environment heavily depend on the application area, this definition is open-ended and needs to be expanded depending on the task at hand. 
For example, a robot vacuum cleaner operates on the floor, where its environment may be defined by the position of walls, furniture, and the presence of dust. 

Agent-based models have been leveraged in real-world settings, including (but not limited to) manufacturing \cite{lim2004integrated}, supply chain \cite{lau2006agent}, transportation \cite{yim2004agent}, and sociology \cite{bianchi2015agent}.
Furthermore, these models are the building blocks of \emph{reinforcement learning}, a paradigm of machine learning research that investigates how agents learn to behave optimally through trial-and-error interactions with their surroundings \cite{sutton1998}. In another example, Lin et at.\ propose a \emph{collaborative} reinforcement learning where agents learn from one another as well as their environment \cite{lin2017collaborative}.
This work extends agent-based models to visual analytic settings where humans and computers interact to conduct analytic tasks.

\subsection{Visual Analytics}

In the early 2000s, a series of unfortunate events in the U.S.\ highlighted the need for more effective utilization of information in decision-making.
These events included the September 11th terrorist attacks and Hurricane Katrina, both of which called for complex analysis of large amounts of information to unravel the ongoing situations and minimize their impacts.
Motivated by this pressing need, Cook and Thomas defined visual analytics as the science of analytic reasoning facilitated by visual interactive interfaces \cite{cook2005illuminating}.
Three years later, Keim et al.\ expanded this definition into the combination of automated analysis techniques with interactive visualizations for an effective understanding of complex datasets \cite{keim2008visual}.
In a nutshell, visual analytics emerged with the goal of forming a human--computer collaboration where each utilizes their respective strengths to solve problems involving data. 

Over the years, researchers have investigated this vast and uncharted visual analytics space from numerous angles. 
For example, some have focused on the complex cognitive processes humans exhibit while making decisions (e.g., \ \cite{kim2021bayesian, wall2017warning, liu2020survey,cashman2019inferential,ottley2020adaptive}), whereas others have focused on improving machine intelligence with the help of humans (e.g.\ \cite{kerrigan2021survey, cabrera2019fairvis,monadjemi2022guided,ha2022unified}). 
In this work, we focus on mixed-initiative visual analytic settings where the system learns from the user's interactions and assists them in their analytic task.

\subsection{Mixed-Initiative Visual Analytics}

Researchers from various backgrounds have investigated the potential and challenges of human--computer collaborations.
They have defined \emph{collaboration} as a process in which two or more agents work together to achieve shared goals \cite{terveen1995overview}.
Furthermore, they have specified \emph{human--computer collaboration} to occur when at least one agent is a human and at least one agent is a computer \cite{terveen1995overview}.
Another term for this collaboration is \emph{mixed-initiative}, where either agent can initiate an action.
Widely accepted in visual analytics settings, Eric Hovitz defines \emph{mixed-initiative user interfaces} as interfaces that enable users and intelligent agents to collaborate efficiently \cite{horvitz1999principles}. 

In this work, we are interested in the pool of mixed-initiative visual analytic settings where humans and intelligent agents collaborate on analytic tasks. 
Specifically, we consider settings in which the role of the computer teammate is beyond just its computational ability (e.g., \ storage and processing of data); rather, we focus on settings where the computer counterpart is an intelligent agent (i.e., learns from users and/or takes actions strategically).
An example of such work is by Sperrle et al.\ \cite{sperrle2021learning}, where they design six artificial agents who guide analysts in refining topic models.

\subsection{Existing Conceptual Models for Visual Analytics}
\label{ss:background_conceptual_models}

Over the years, visual analytics researchers have faced a set of theoretical questions such as \emph{what is the purpose of visual analytics} \cite{andrienko2018viewing}, \emph{how do humans generate knowledge} \cite{sacha2014knowledge}, and \emph{what are the strengths of humans and computers in collaborative data analysis} \cite{crouser2012affordance}. 
They have offered their answers to these questions as a set of \emph{conceptual frameworks} or \emph{conceptual models}, which have evolved over time. 
Here, we provide a brief summary of some of these models and their evolution.

One of the earliest conceptual models widely adopted in visual analytics is the \emph{sense-making loop} by Pirolli and Card, which describes \emph{how analysts make sense of data} \cite{pirolli2005sensemaking}. Their model presents sense-making as a process that involves searching for relevant information (i.e., \ foraging) and generating/confirming hypotheses (i.e., \ synthesis). Their model primarily describes how \emph{human agents} operate to understand data.

As the visualization field began to mature, van Wijk aimed to understand \emph{the purpose and meaning of visualizations} \cite{van2006views}. They proposed an economic model of data visualization in which the value of visualization is defined as the insight it provides. Furthermore, the user is assumed to iteratively refine the visualization specifications in the exploration process. Years later, Ceneda et al.\ augmented this model to include \emph{guidance}, which they define as a \emph{computer-aided process that actively resolves user knowledge gaps in interactive visual analytics} \cite{ceneda2017chracterizing}. Recently, Sperrle et al.\ developed \emph{Lotse}, a library that bridges the gap between theory and practice for visual analytic guidance \cite{sperrle2022lotse}.

Focusing on the promise of visual analytics in uniting humans and computers, Keim et al.\ envisioned a tight integration of visual and automated analysis methods \cite{keim2008visual}. Sacha et al.\ expanded on how the visual analytic process contributes to knowledge generation \cite{sacha2014knowledge}. In a recent development, Andrienko et al.\ further expanded on this model, arguing that the purpose of visual analytics is to build a model of some piece of reality \cite{andrienko2018viewing}.

We characterize the existing conceptual frameworks of \acro{VA} in Table \ref{tab:unifying_existing} along three dimensions: whether they describe human agents (e.g., \ analysts), whether they describe the visual analytic environment (e.g., \ data, visual interface, etc.), and whether they describe artificial agents (e.g., \ models, automated techniques, etc.). We will expand on each of these dimensions further in Section \ref{sec:agent-based-framework}.

\begin{table}[h]
\centering
\footnotesize
\caption{Existing conceptual models in visual analytics, organized by whether they describe human agents \hagenticon, the visual analytic environment \envicon, or artificial agents \cagenticon.}
\label{tab:unifying_existing}
\begin{tabularx}{\linewidth}{@{\extracolsep{\stretch{1}}}{l}*{3}{c}@{}}
\toprule
Existing Conceptual Models &  \hagenticon & \envicon & \cagenticon \\ 
\midrule
Sensemaking Loop \cite{pirolli2005sensemaking} & \faCheck & & \\
Views on Visualization \cite{van2006views} & \faCheck & \faCheck & \\
Multi-analyst Collaborative Framework \cite{brennan2006toward} & \faCheck & \faCheck & \\
Visualization Exploration \cite{tjvisualizationexploration} & \faCheck & \faCheck & \\
Visual Analytic Process \cite{keim2008visual} & \faCheck & \faCheck & \faCheck \\
Affordance-based Framework \cite{crouser2012affordance} & \faCheck & \faCheck & \faCheck \\
Knowledge Generation Model \cite{sacha2014knowledge} & \faCheck & \faCheck &  \faCheck \\
Guidance in Visual Analytics \cite{ceneda2017chracterizing} & \faCheck & \faCheck & \faCheck \\
Visual Analytics as Model Building \cite{andrienko2018viewing} & \faCheck & \faCheck & \faCheck \\
\bottomrule
\end{tabularx}
\end{table}

\section{Research Goals}

The existing conceptual models outlined in the last section provide us with an understanding of analysts' workflows, visual analytic systems, and approaches for guidance. 
As visual analytic settings evolve into a more sophisticated human--computer collaboration, the existing frameworks lack the vocabulary to describe this team dynamic.
To address this gap, we introduce an agent-based framework for \acro{VA} which is drawn from the \acro{AI} literature. Using this framework, we aim to:

\begin{enumerate}
    \item[G1] Describe the full visual analytic pipeline under the agent-based model, defining analysts and visual analytic settings from the lens of \emph{human agents} and \emph{environments}, respectively;
    \item[G2] Characterize high-level analytic behavior according to existing classifications of intelligent agents in the \acro{AI} literature, providing a balanced abstraction in better understanding analysts; and 
    \item[G3] Highlight opportunities for analytic guidance to be provided by \emph{artificial agents}, where effective guidance relies on understanding the analysts and mapping them to appropriate \acro{AI} teammates. 
\end{enumerate}

\section{The Agent-based Framework for Visual Analytics}
\label{sec:agent-based-framework}

\begin{figure*}[!t]
    \centering
    \includegraphics[width=\textwidth]{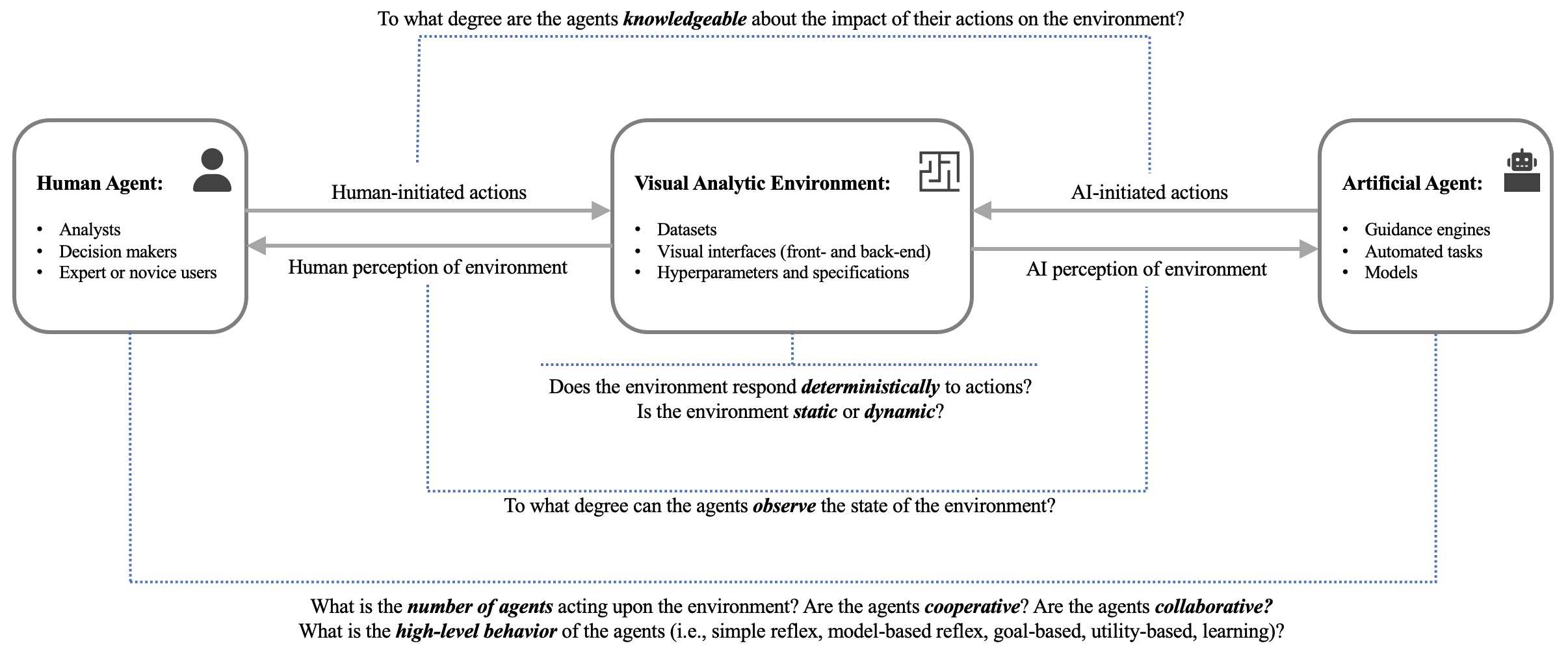}
    \caption{An agent-based framework for visual analytics and its attributes.}
    \label{fig:agent-va-framework}
\end{figure*}

The overarching goals of visual analytics research are to better understand the analytic workflows and to improve them by leveraging automated techniques alongside human strengths. As we move towards more complex visual analytic settings involving multiple analysts and automated techniques collaborating, we need a conceptual framework to reason about the entire system.
In this section, we argue that the visual analytic process is synonymous with agents interacting with their environment (Fig.\ \ref{fig:agent-va-framework}).
For example, analysts interact with visual interfaces and complex datasets to generate hypotheses, answer questions, and make decisions.
In order to view visual analytics from this agent-based perspective, we need to specify \emph{agents} and \emph{environments} in this context and discuss their key attributes. 
In doing so, we use the following two hypothetical scenarios as running examples throughout this section.

\noindent \textbf{Scenario 1:} Alice is a data scientist who is developing a deep-learning model for detecting anomalous flights in the national airspace. She has utilized her expertise in machine learning to design the model architecture. Once the model is trained and optimized using historical flight data, the performance metrics are visualized on an interactive interface. Furthermore, given the high dimensionality of her data and various performance measures, the system automatically recommends visualizations that may be insightful in diagnosing the model.

\noindent \textbf{Scenario 2:} Bob is a novice user who is using a popular real estate website (containing an interactive map visualization) to find his next house to purchase. Given that he is new to the city, he is overwhelmed by a large number of houses for sale and his lack of experience with the local real estate market. As he browses the houses and bookmarks his favorite ones, an algorithm passively learns about his latent interests and highlights the houses on the map in which he may be interested.

\subsection{Visual Analytic Agents}

Visual analytic agents are entities capable of observing and acting in order to contribute to an analytic objective. As we rapidly move towards mixed-initiative visual analytic settings, such entities may be \emph{human} or \emph{artificial} agents. 

Human agents consist of data scientists, decision-makers, domain experts, novice users, or any other user group who may conduct analytic tasks using data. Prior work has aimed to understand the needs of different user groups (e.g.,\ \cite{wong2018towards}) and investigate how individual differences may impact their workflows \cite{xu2020survey}. Furthermore, they have investigated how humans perceive data (e.g.,\ \cite{xiong2022seeing}) and what kinds of actions they take during analytic sessions (e.g.,\ \cite{gotz2009characterizing, brehmer2013multi}).

Artificial agents, on the other hand, consist of modeling algorithms, guidance engines, and automated tasks which interact with the environment to contribute to a shared analytic task. Prior work has attempted to design artificial agents who uncover patterns in the data (e.g.\ \cite{kim2019topicsifter}), learn from user interactions (e.g.,\ \cite{brown2012dis}), and guide the user during their session (e.g.,\ \cite{dabek2016grammar}).

With these descriptions of human and artificial agents in the context of visual analytic settings, we next identify the agents present in each of our running examples.

\noindent \textbf{Agents in Scenario 1: } There are one human agent and two artificial agents in this scenario. Alice is the human agent who is interested in developing a model. The deep learning and visualization recommendation algorithms are the two artificial agents that intelligently take actions according to their observations (i.e.\ tune parameters to learn from historic data and select visualizations to recommend). The common objective is to develop and diagnose a model that accurately identifies anomalous flights in the airspace.

\noindent \textbf{Agents in Scenario 2: } There are one human agent and one artificial agent in this scenario. Bob is the human agent who is searching for his next house. The user modeling algorithm is the artificial agent which is aiming to learn Bob's latent interests and visually prioritize houses that may cater to his interests. The common objective is to identify Bob's favorite house.

\subsection{Visual Analytic Environments}

Visual analytic environments are specified by the collection of datasets, visual interfaces, and configurations with which the agents (both human and artificial) interact. 
Datasets are external sources of information from a particular domain that are typically stored in databases. 
Visual interfaces consist of the front- and back-end programs that provide the user with intuitive means of perceiving and interacting with the information effectively.
Configurations refer to any settings, specifications, or hyper-parameters that govern various aspects of the analysis.

Agents may interact with any of these components and observe the outcome of their actions.
They may interact with datasets by aggregating, transforming, or wrangling the information.
They may interact with visualizations by brushing, hovering, or clicking data points.
They may interact with configurations by modifying the mapping of data features to visual channels in order to generate more insightful visualizations or modifying their choice of the distance function which in turn updates the underlying models. 
These are just some examples of how agents may interact with each of the components in the visual analytic environment.

Consider applying this definition of visual analytic environments to the running examples.

\noindent \textbf{Environment in Scenario 1: } The objects in this environment include the historic flight database, the visualizations for inspecting the data and model performance, and the specifications for the model architecture (e.g. number of hidden layers, activation functions, etc.).

\noindent \textbf{Environment in Scenario 2: } The objects in this environment include the real estate database, the visualization interface, and the filters the user has applied.

\subsection{Attributes of Agents and Environments} 
\label{sec:agent-based-attributes}

\acro{AI} researchers have proposed a set of dimensions by which they characterize agents and environments \cite{russellnorvigaibook}. 
These dimensions have guided innovations in designing artificial agents appropriate for given tasks, and have been used to categorize these innovations into focus areas of \acro{AI}.
Below, we discuss some of these dimensions in the context of visual analytic settings.

\emph{Observability } refers to the degree by which an agent can perceive the environment. In visual analytic settings, observability could be determined by both the environment and the agents' abilities. For example, having missing and uncertain data makes our visual analytic environment less observable. As another example, human factors such as limited cognition and perception may make aspects of the environment less observable.

\emph{Knowledge } refers to the agents' level of understanding towards the ``laws of physics'' in the environment. We say an environment is \emph{known} (\emph{unknown}) to the agent if the agent knows (does not know) how their actions impact the environment. In the visual analytic settings, this attribute maps to analysts' expertise, experience, and familiarity (with visualizations, domain, analysis, interfaces, etc.). For example, an experienced machine learning engineer can reason about how tweaking certain parameters will impact the results (the environment is known to them). In contrast, a less experienced engineer may tweak parameters by chance in an attempt to learn the inner workings of the models and improve their performance.

\emph{Number of agents } who act upon an environment can vary depending on the application. For example, one vacuum robot cleaning the floor is considered a \emph{single-agent} system, whereas multiple cars driving on the highway constitute a \emph{multi-agent system}. Similarly, we may have one or more analysts attempting to answer questions in visual analytic settings. Furthermore, we may have artificial agents (e.g.\ guidance engines) providing help to human analysts.

 \emph{Level of cooperation and collaboration } refers to the degree to which agents share rewards and goals. In fully cooperative settings, agents aim to achieve the best societal outcome (i.e.\ the outcome that provides the most collective benefit) whereas, in less cooperative settings, agents focus on individual outcomes (i.e.\ the outcome that benefits the agent most individually). Furthermore, in collaborative settings, agents work toward a common goal whereas, in adversarial settings, agents work against one another. We believe that most visual analytic settings to date are both cooperative and collaborative, as the goal for the human and the computer is to improve the analytic process.

 \emph{Dynamics } refers to the degree by which an environment changes over time. On one extreme, we have static environments that do not change, and on the other extreme, we have dynamic environments that change rapidly. In visual analytic settings, this characteristic can refer to the data changing, a model evolving, or an interface adapting over time. For example, analysts who monitor elections often work with dynamic datasets that evolve as new polls close and new results are reported. 

 \emph{Determinism } refers to the level of certainty in the outcome of an action. In deterministic environments, an action given a fixed state will always result in the same outcome whereas in stochastic environments, an action given a fixed state may result in different outcomes following a probability distribution. An example of this in visual analytic settings is that performing dimensionality reduction on the same dataset may result in different views due to some randomness in the underlying processes.

Next, we characterize the running examples using these attributes. We note that a full characterization may require more details about the scenario and may be subjective. Our goal is to offer one possible characterization of the scenario for demonstration.

\noindent \textbf{Attributes of Scenario 1: }  
We claim that the model is fully observable to Alice since she can see the impact of her design decisions on the model through the performance metrics and visualizations. Due to the large amount and potentially noisy nature of historical flight data, we characterize the data to be only partially observable to Alice. Alice is knowledgeable in model building due to her expertise in data science, however, she has limited domain knowledge towards aeronautics. This is an example of a multi-agent setting (one human and two \acro{AI} agents). The agents work cooperatively and collaboratively to build a high-quality model. Assuming the historic flight data does not change, this system is static. Assuming the optimization engine and dimensionality reduction techniques are randomized, then we can consider this system to be stochastic.

\noindent \textbf{Attributes of Scenario 2: }
In this example, the dataset is fully observable to Bob since he can select any house and see more details about it. Since Bob is new to the city, he has limited knowledge towards the real estate market in the city. 
This is an example of a multi-agent setting (one human and one \acro{AI}).
The agents here work collaboratively and cooperatively. If the \acro{AI} agent was to highlight houses in exchange for a premium paid by sellers, then we could consider this setting to involve a conflict of interest (i.e.\ the \acro{AI} trying to help sellers instead of Bob). Since the housing market can change in real-time by new houses being added, some houses going under contract, and prices changing, we characterize this example to be a dynamic system. Assuming there is no randomization involved, this is a deterministic system.

\subsection{Are Agents Part of the Environment?}

When thinking about visual analytic settings using this agent-based framework, we anticipate researchers and practitioners wondering if agents are also part of the environment. We suggest that the answer to this question depends on the perspective from which we observe the system.

\begin{figure}[!ht]
    \centering
    \includegraphics[width=\linewidth]{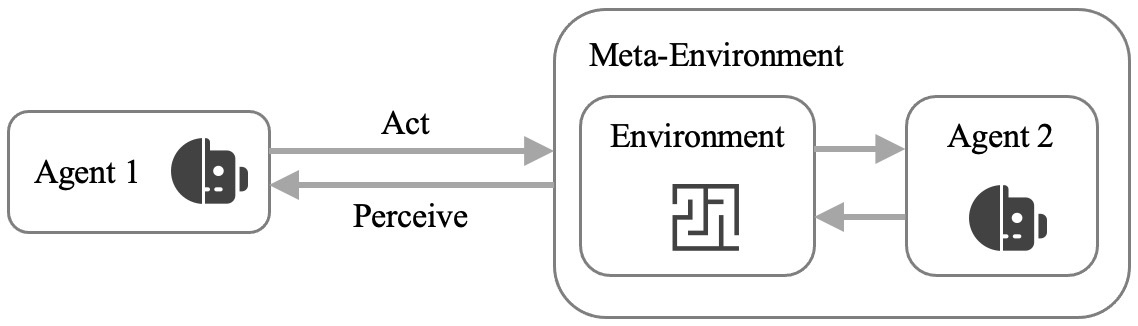}
    \caption{The nested structure of the environment which Agent 1 perceives and acts upon.}
    \label{fig:nested-agent}
\end{figure}

From a third-party perspective, the system looks as shown in Figure \ref{fig:agent-va-framework}, where agents are not part of the environment (but they interact with the environment).
However, from each individual agent's perspective, the system can be viewed as a \emph{me vs.\ them} setting where the individual agent considers other agents and the environment to all be part of a meta-environment with which it interacts (Figure \ref{fig:nested-agent}).
This is in-line with our observations in the visual analytics literature where agents do not directly interact with one another. Instead, they interact with one another through the environment (e.g.\ the visual interface). In Figure \ref{fig:nested-agent}, for example, Agent 1 learns about Agent 2 by observing its interactions with the environment. Furthermore, Agent 1 may interact with Agent 2 indirectly by manipulating the the environment.

\section{Types of Behavior in Visual Analytic Agents}
\label{sec:analytic-behavior}

In the previous section, we presented agents, environments, and some of their attributes in the context of visual analytics. 
In this section, we utilize those definitions to characterize user behavior similar to how \acro{AI} researchers have characterized artificial agent behavior \cite{russellnorvigaibook}.
We assert that these categorizations are not mutually exclusive. For example, a user may seek a desired end state (goal-based) but may select actions to minimize the duration of their pursuit (utility-based). Similarly, analysts may begin their session by familiarizing themselves with the data (learning) and switch their intents as the tasks evolve. Throughout this section, we refer to Scenarios 1 and 2 from Section \ref{sec:agent-based-framework} as running examples.

\subsection{Simple Reflex Agents}

In the most simple case, we assume that agents are able to fully observe the current state of the world and are knowledgeable about how their actions will impact the world. Therefore, the simple behavior of agents can be characterized as a set of if-else statements, also referred to as \emph{simple reflex}:
\begin{quote}
    Simple reflex agents choose their actions based on pre-defined rules given the current state of the world. 
\end{quote}

An example of simple reflex behavior in human analysts is interactive data labeling settings. These examples may utilize humans' perception to quickly identify certain objects and share that information with the underlying algorithms as labels \cite{bernard2017comparing}. In Scenario 2, for example, the user may save houses that meet a certain set of rules or characteristics as an initial step to narrow down the search space.

\subsection{Model-based Reflex Agents}

When agents do not know the current state of the world (i.e.\ the environment is not fully observable), then they try to maintain a mental model of the state of the world based on their experience. These agents will then utilize their mental model of the world to operate via if-else statements, a behavior referred to as \emph{model-based simplex reflex}:

\begin{quote}
    Model-based reflex agents choose their actions based on pre-defined rules given their guess (or mental model) towards the current state of the world.
\end{quote}

An example of a model-based reflex behavior in human analysts is when they aim to make decisions based on uncertain, missing, and inconsistent datasets. Hence, they use their best judgment toward the true state of the world to proceed with decisions. This behavior may suffer from issues such as analyst bias. 
In Scenario 1, for example, Alice is not as familiar with aeronautics data. Therefore, she may attempt to clean and filter the data based on her best guess of how a set of filtering criteria may impact the data.

\subsection{Goal-based Agents}

In some cases, the agents have a goal state they pursue in addition to being aware of the current state of the world. These agents aim to find a sequence of actions that takes them from their current state to the desired goal state. This behavior is referred to as \emph{goal-based}:

\begin{quote}
    Goal-based agents choose the action that gets them closer to their goal state, given the current state of the world. 
\end{quote}

This is markedly similar to assumptions of existing visualization research and task models~\cite{lam2017bridging,battle2019characterizing}.
In analytic scenarios, we often assume that the analyst has a particular objective that drives their tasks and actions and explores until they either uncover evidence that aligns with their initial objective or they refine or update their hypotheses and goals. 
Assuming that the analyst's actions are goal-based, it may be possible to apply AI algorithms to model behavior and predict their interactions and goal state. For example, search and planning are two AI subfields dedicated to learning an agent's sequence of actions toward their destination. 
In Scenario 1, for example, one of Alice's goals is to clean and prepare the data for her machine learning algorithm. In doing so, she will take a sequence of filtering steps until the data meets her desired quality.

\subsection{Utility-based Agents}

A goal-based agent can only make a binary decision about the desirability of the state based on whether or not it's a goal state.
In contrast, \emph{utility-based} agents use a utility function to reason about the desirability of the resulting state from each action:

\begin{quote}
    Utility-based agents can compare actions according to a utility function quantifying its desirability, and select one that results in the highest expected utility.
\end{quote}
 
For example, the information foraging theory states analysts' desire to maximize information intake while minimizing effort \cite{pirolli1999information}. 
This example applies to visual analytic settings where analysts wish to identify as many relevant points as possible while minimizing distractions with irrelevant points \cite{monadjemi2022guided}.
In Scenario 2, for example, Bob has a mental utility function which ranks some houses higher than others based on a combination of their characteristics. By optimizing his utility function, Bob will ideally only visit the most promising houses in person.

\subsection{Learning Agents}

Thus far, we have assumed that agents know the ``laws of physics'' governing their environment. In other words, we have assumed that agents know how a given action impacts the state of the world. In contrast, when agents operate in an unfamiliar environment and improve as they gain experience, they exhibit the \emph{learning} behavior:
\begin{quote}
    Learning agents start with incomplete knowledge of their environment and become more knowledgeable as they gain experience through interactions. 
\end{quote}

Consider exploratory visual analysis as an example. Analysts begin with forming questions they wish to answer with data (i.e., \ lack of knowledge towards a phenomenon) and iteratively update their understanding as they explore the data \cite{battle2019characterizing}. 
In Scenario 1, for example, Alice iteratively updates her understanding of aeronautics data sources as she explores the dataset and applies various transformations to the data.

\section{Case Studies}

Using three examples from the literature, we now demonstrate how our framework can be applied to characterize visual analytic settings.
Upon conducting an extensive literature review of mixed-initiative \acro{VA} work in the past two decades, we selected these examples because they included detailed descriptions of the tasks users were asked to perform, as this level of detail is necessary to characterize user behavior and the analytic setting. 
For each example, we begin with a summary of the scenario, then characterize the visual analytic environment, human agents, and artificial agents.

While characterizing human agents' analytic behavior according to the classifications in Section \ref{sec:analytic-behavior}, we uphold our philosophy that an analyst may not exclusively fit into only one class. 
Rather, they may exhibit any of the behaviors simultaneously or switch between them sequentially. Therefore, instead of trying to fit users in each classification, we will discuss \emph{how} the users may exhibit some of the outlined analytic behaviors in their respective sessions.

\subsection{Finding Waldo}
\label{ss:example-waldo}

\begin{figure}[!h]
    \centering
    \includegraphics[width=0.9\linewidth]{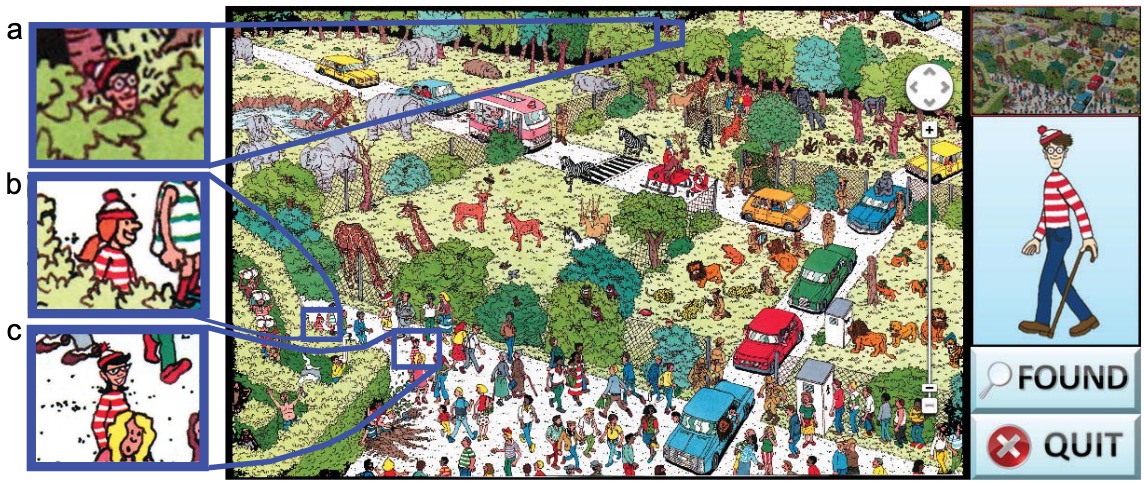}
    \caption{The interface used in the Finding Waldo experiment \cite{brown2014finding}. (a) is the target named Waldo, (b) and (c) are distractors who are dressed similarly to Waldo.}
    \label{fig:waldo-interface}
\end{figure}

Brown et al.\ \cite{brown2014finding} investigated if analysts' interaction log can uncover their personality traits.
They asked users to perform a visual search task, namely, to find Waldo.
In this well-known and challenging task, users were presented with a high-resolution image of a crowded garden and were asked to find a cartoon character named Waldo. 
To mimic some qualities of real-world scenarios, several distracting characters were dressed similarly to Waldo. However, they were not the target (Figure \ref{fig:waldo-interface}).

\subsubsection{Environment}

In this example, the visual analytic environment consists of the image data and the interactive interface. 
The interface offers a set of affordances for navigating the image (up, down, left, right, zoom-in, zoom-out). 
In response to calls to each functionality, the viewpoint is updated to show the respective portions of the image.

Using the attributes discussed in Section \ref{sec:agent-based-attributes}, we characterize this visual analytic setting as follows. The state of the environment is observable at various degrees depending on the zoom level. While there are no missing data, the user can perceive limited details at zoomed-out views. Since the user is fully aware of the impact of different interactions, this environment is known. Lastly, we claim that this setting is single-agent, static, and deterministic since there is only one agent operating, the image is static (i.e.\ chatacters do not move), and the impact of an action from a given state is fixed.

\subsubsection{Human Agents}

This setting involves one human agent interacting with the interface to find Waldo. Once found, the user has accomplished their goal and presses the FOUND button to conclude the session.

We argue that the primary analytic behaviors exhibited by users in this task are \emph{goal-based} and \emph{model-based reflex}. 
Throughout the entire session, the user works towards a goal state: to identify Waldo's position in the image (shown in Figure \ref{fig:waldo-interface}-a).
Due to the perceptual limitations at the zoomed-out view, the user may only be able to guess where Waldo might be. For example, they may choose to explore areas with red since they know Waldo's shirt is also red.
This maintained belief towards Waldo's position makes the user exhibit behaviors of a model-based reflex agent.

While we consider the behaviors above to be the primary ones for this task, one could argue that the user exhibits aspects of other kinds of agents as well. For example, with every individual the user observes in the garden, they try to match it with Waldo's known appearance and perform a set of if-else statements to decide if the individual under consideration is in fact, Waldo. This behavior is similar to one by simple reflex agents. Furthermore, one could argue that the user is not only trying to locate Waldo, but rather, the user wants to locate Waldo as quickly as possible. This would make the user a utility-based agent trying to optimize for time spent searching (or any other arbitrary utility function).

\subsubsection{Artificial Agents}

By the definition that agents \emph{perceive} and \emph{act}, we claim that this visual analytic setting does not involve artificial agents.
Rather, the authors collected interaction data and analyzed them after the session to predict task completion speed and its relationship with personality traits. 
In a sense, there is an aspiration to design artificial agents which recognize users' personality traits and provide them with appropriate guidance.
In an extension, we envision this work being augmented with state-of-the-art goal recognition techniques (e.g.\ \cite{shvo2020active, van2021activity}) to identify users' goal state and assist them during the discovery process while taking into consideration human factors such as personality traits.

\subsection{Reducing Visualization Latency with ForeCache}

Battle et al.\ \cite{battle2016dynamic} investigated visualization latency caused by database queries. Such latency is known to adversely impact user experience and hinder data exploration \cite{liu2014effects}. 
To alleviate some of the latency, they proposed maintaining a model over user interactions to anticipate future actions and pre-fetch the data proactively. 
For evaluation, the authors designed a map-based visualization  of the \acro{NASA MODIS} snowfall data across America. Users were asked to explore the data and identify specific regions with the highest amount of snowfall. 
Due to the high resolution, it would not be feasible to visualize the entire dataset. Therefore, the system aggregated the data into lower-resolution tiles for overview and increased the granularity as users zoomed in.

\begin{figure}[!h]
    \centering
    \includegraphics[width=0.8\linewidth]{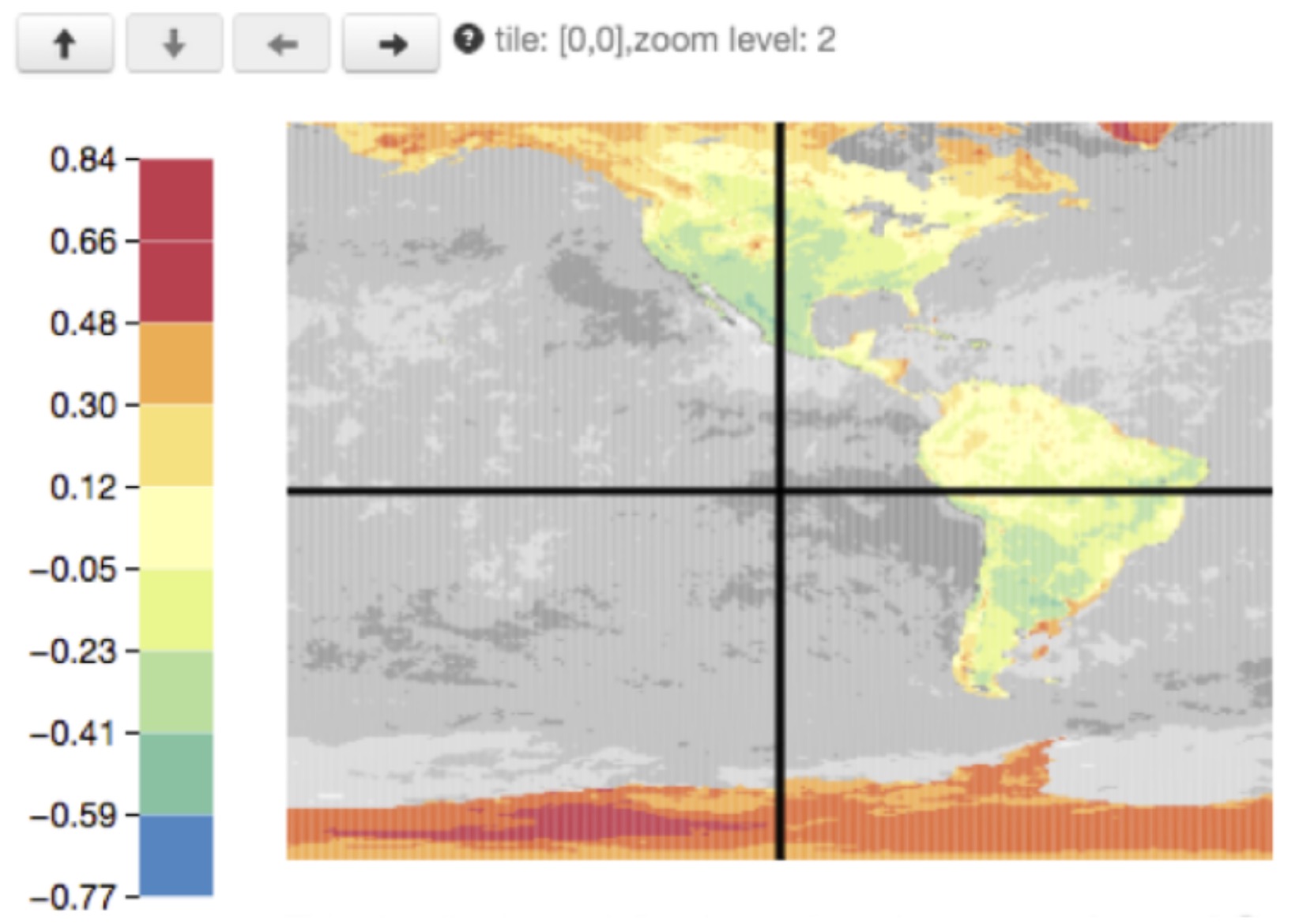}
    \caption{The interface used in the ForeCache project, visualizing the amount of snow recorded in the \acro{NASA MODIS} dataset \cite{battle2016dynamic}.}
    \label{fig:forecache-interface}
\end{figure}

\subsubsection{Environment}

The \acro{VA} environment consists of the \acro{NASA MODIS} dataset, the interactive map visualization (front-end), and the data storage/processing infrastructure (back-end). 
The interface offered a set of affordances for panning and zooming similar to the last example. 
Each interaction updated the user's viewpoint according to pre-defined rules. 
As users zoomed in, the system increased the resolution of the visualization for a more detailed view. As users zoomed out, the system would aggregate the data into lower-resolution tiles. 
Using the attributes outlined in Section \ref{sec:agent-based-attributes}, we characterize this visual analytic setting to be very similar to the one in Section \ref{ss:example-waldo}: varying degrees of observability depending on zoom level, known, static, and deterministic. We omit justifications for these choices as they are identical to our last example. The main difference between the two environments is that this environment is multi-agent since there is one human and one artificial agent.

\subsubsection{Human Agents}

The human agent in this example was a domain scientist who would interact with the environment via the front-end controls in order to identify locations with the highest snowfall. 
Using the behavior types discussed in Section \ref{sec:analytic-behavior}, we argue that the user performing this task primarily exhibits \emph{utility-based} and \emph{model-based} analytic behaviors. 
In this case, the utility function to be maximized may be defined by the amount of snowfall and the number of discovered tiles.
Since the users are domain experts, we expect them to have a mental model of the data. For example, they may expect more snow in Colorado than in Texas, hence exploring Colorado more.

\subsubsection{Artificial Agents}

In contrast to our last example, this example involves an artificial agent observing and acting on the environment.
The artificial agent is the ForeCache algorithm which observes user interactions and initiates data queries in anticipation of future interactions. 
We note that this is an example where the artificial agent does not take action on the front-end. 
Rather, it takes actions in the back-end which are not visible to the user. 
Nonetheless, the artificial agent impacts user experience by reducing latency.

We characterize this artificial agent as a \emph{model-based reflex} and \emph{learning} agent. This agent does not have access to future information on how the user will navigate, therefore, it maintains a Markov chain model that decides the most likely outcome only based on the user's current view (hence the term \emph{Markov}). This model, however, was not given to the agent a priori. By observing user interactions, the agent maintains and updates its belief over how the user may act given its current state, making the agent a learning agent.

\subsection{Optimizing Furniture Delivery Schedules}

\begin{figure}[!h]
    \centering
    \includegraphics[width=0.9\linewidth]{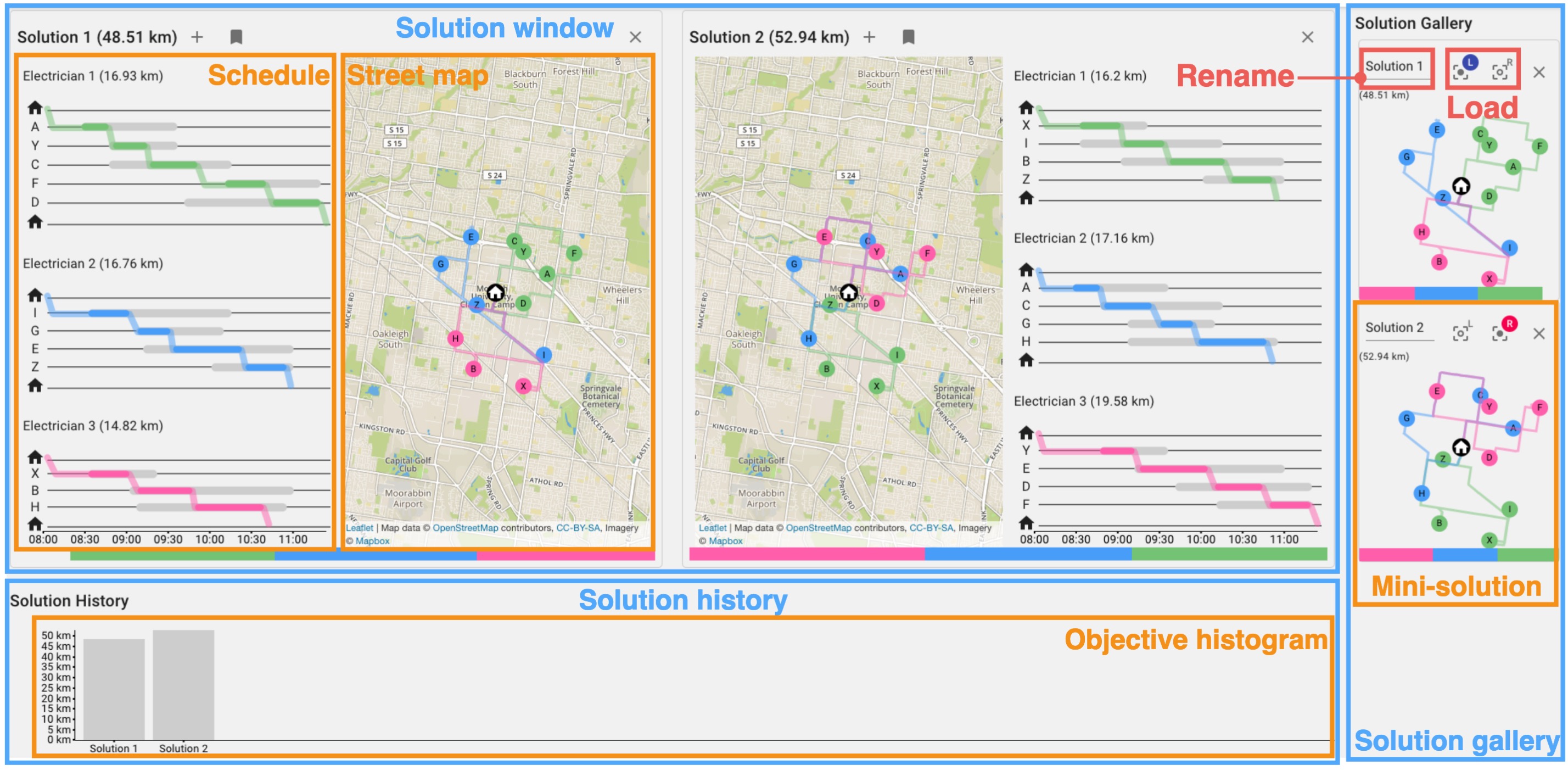}
    \caption{The interface used for optimizing the furniture delivery schedule subject to customer and supplier constraints \cite{liu2020supporting}.}
    \label{fig:scheduling-interface}
\end{figure}

Liu et al.\ \cite{liu2020supporting} consider a different problem than our previous examples: solving an optimization problem subject to a set of constraints. 
They present a hypothetical situation where a home furnishing store receives orders from their customer and needs to assign each order to a delivery truck. Deliveries must be made during customer availability windows, and each delivery may take a different amount of time. 
The furnishing store would like to identify an order-to-truck assignment that ensures every order is delivered at an appropriate time and the total distance traveled by the trucks is minimized.
Hence, they design a mixed-initiative system for scheduling their deliveries. The system considers the set of known constraints and proposes a set of solutions. The user can inspect the solutions and modify them as needed to incorporate other constraints that may be missing from the knowledge base (e.g.\ emergencies with trucks being out of service or customers having special requests). Figure \ref{fig:scheduling-interface} shows the interface used for this experiment.

\subsubsection{Environment}

In this example, the visual analytic environment consists of the set of customer orders, destination addresses, customer availability windows, trucks available for service, and the interactive interface. 
The interface contains visualizations of solutions (both timeline and map view), a history of solutions considered during a session, the total mileage driven for each solution, and the gallery of saved solutions by the user. 
In terms of interaction design, this interface provides affordances for \emph{exploring the solutions} (bookmark, rename, modify, re-optimize, load) and \emph{adding constraints} (assign deliveries to requested trucks, rearrange the order of deliveries).

Using the attributes discussed in Section \ref{sec:agent-based-attributes}, we characterize this visual analytic setting as follows. This environment is fully observable to the human agent and partially observable to the artificial agent. This is because the human agent has access to all the information, whereas the artificial agent is unaware of some customer availability constraints. We consider this environment to be known to the agents since they are aware of the impact of their actions on the environment. We assume that the orders and constraints for a given day are fixed, making the environment static. This would be a dynamic setting if orders were to be constantly added and canceled for same-day delivery. Since there are two agents operating in this environment (the human and the optimization algorithm), this is a multi-agent setting. The interactions between the interface and the user are deterministic, however, the underlying optimization algorithm may be stochastic.

\subsubsection{Human Agents}
The human agent in this example is an end user who interacts with the interface in order to schedule the deliveries.
Using the behavior types discussed in Section \ref{sec:analytic-behavior}, we characterize them to exhibit \emph{goal-based} and \emph{utility-based} behaviors. 
While the goal driving interactions is to create a schedule subject to constraints, the user also may have a utility function to optimize. In this example, that utility function was defined by the total number of miles driven by the trucks. However, the user could be interested in other measures of utility such as workload balance or the chance of delays.

\subsubsection{Artificial Agents}
The artificial agent is the optimization algorithm, aiming to find viable delivery solutions that satisfy the constraints and optimize for the total distance traveled. 
To do so, this agent utilizes Constraint Satisfaction tools such as MiniZinc \cite{nethercote2007minizinc} and Gecode \cite{gecode}.
Similar to the human agent, we argue that the artificial agent is also primarily exhibiting \emph{goal-based} and \emph{utility-based} behaviors.

\section{Discussion and Future Work}
In this section, we reflect on the current state of mixed-initiative \acro{VA} research in light of our framework and case studies. Furthermore, we discuss the implication of our framework on the design of future \acro{VA} systems as well as open questions for future work.

\subsection{State of the Art from the Agent-based Perspective}

We described the agent-based framework as one with the potential to unify the entire \acro{VA} pipeline.
Using this framework, we believe any \acro{VA} setting can be characterized as a subset of humans and \acro{AI} teammates interacting with their environments to solve problems involving data.
It is our hope that this framework will help researchers maintain a high-level view on how various sub-fields of visual analytics research fit into the broader puzzle.
While we highlighted only three examples as case studies, we emphasize that we can reason about many more instances using this framework.
For example, the work by Xiong et al.\ \cite{xiong2022seeing} investigates how humans perceive correlations on scatter plots. This can be characterized as a study focused on the arrow from the visual analytic environment to the human agent in Figure \ref{fig:agent-va-framework} (i.e. how human agents perceive the environment). 
As another example, existing interaction taxonomies (e.g.\ \cite{gotz2009characterizing, amar2005low, yi2007toward}) may be viewed as a description of how human analysts interact with visual analytic environments.
As we saw in our first case study, our framework can highlight some gaps in the literature as well. The work by Brown et al.\ \cite{brown2014finding} utilizes interaction logs to predict the user's personality traits. This can be characterized as a study focused on the arrow from the environment to the artificial agent in Figure \ref{fig:agent-va-framework} (i.e. how can the \acro{AI} teammate perceive and reason about other agents). 
However, our framework highlights the gap to investigate how an artificial agent may \emph{act} in light of observing the users' personality traits to contribute to the task effectively.

\subsection{Designing Future Visual Analytic Systems}

In addition to reasoning about existing \acro{VA} settings, this agent-based framework can facilitate the design of future \acro{VA} systems. In doing so, researchers and practitioners may begin by specifying various components of the environment, identify the agents, and design artificial agents accordingly to contribute to the analytic task. As an example, we refer to the work by Sperrle et al.\ \cite{sperrle2021learning} which designs a multi-agent visual analytic system with one human agent and six \acro{AI} agents for refining topic models.

\subsection{On Characterizing High-level Human Behavior}

In the case studies, we aimed to characterize analysts' high-level behavior by drawing inspiration from how \acro{AI} researchers characterize artificial agents' behaviors.
We described some agents as ones who take actions to fulfill a goal and others who may take actions based on their mental models of the world (among others).
While considering a large pool of papers from the literature and attempting to characterize agent behavior, however, we found the process to be more ambiguous for some papers than others. 
Upon some reflection, we noticed that papers that present detailed descriptions of the analytic task in their user studies were easier to characterize. 
This challenge poses the question of how much information is necessary in order to fully characterize high-level behavior using our framework.
Furthermore, we noticed that agent behavior may change or evolve over time \cite{sperrle2021co}. Therefore, the characterization of agent behavior should realistically be a time-series data, where users may exhibit different behaviors at each timestep. 
We hope that our work is a first step towards mapping analytic sessions to a sequence of high-level behaviors in order to move towards describing the analytic intent behind \emph{why} analysts take certain actions (as opposed to \emph{what} actions they take).

\subsection{Towards Learning Analytic Intent as Utility Functions}

Under the assumption that analysts are sometimes utility-based agents, then we can pose a question on how to learn and represent analytic intent as a mathematical utility function. 
\acro{AI} researchers have explored similar ideas under the umbrella of \emph{active learning} algorithms. Active learning algorithms are designed to acquire data labels strategically in order to fulfill an analytic objective. For example, the uncertainty sampling algorithm \cite{settles2009active} defines utility in terms of the amount of information learned by making an observation, hence the algorithm selects data points with the highest uncertainty in their label. Another example is the active search algorithm \cite{jiang2017efficient} which defines utility as the total number of relevant data points discovered by the end of a session, where members of a given class are deemed to be relevant. Active search has recently been augmented into visual analytic workflows for data foraging \cite{monadjemi2022guided}. 
Recently, \acro{VA} researchers have investigated utility functions from human-centric and data-centric perspectives. In the human-centric approach, the utility functions are elicited from the users directly \cite{10.2312:eurova.20221072}, where they can rank their preferences along various data dimensions to create a utility function. 
In the data-centric approach, on the other hand, Bernard et al.\ \cite{bernard2021taxonomy} present a comprehensive set of data characteristics that may be used for defining utility functions.
We envision more investigations in this area on different forms of human-centric utility functions, their value to the analytic tasks, and seamless methods of eliciting them.
This could be another step towards building more effective artificial agents for mixed-initiative visual analytic settings.

\subsection{Limitations}

We proposed this agent-based framework with the goal of further bridging the gap between \acro{VA} and \acro{AI}. 
We envisioned for this framework to help us organize \acro{VA} research, characterize human behavior, and design more effective artificial agents for analytic tasks.
We do, however, acknowledge that there are limitations.
For example, the list of analytic behaviors from Section \ref{sec:analytic-behavior} should not be taken as a comprehensive list. 
While that characterization of agents is well established and taught in \acro{AI} classrooms often, they also evolve over time to include cases not considered before.
For example, recent work in \acro{AI} has focused on artificial agents who learn from their mistakes \cite{zhi2020online}, those who are able to solve problems creatively \cite{gizzi2022creative}, and those who learn from other agents \cite{lin2017collaborative}.
Another limitation to consider is the ambiguity in some of the definitions. 
This limitation is one that is inherent to the agent-based models. 
As discussed by Wooldridge and Jennings \cite{wooldridge1995intelligent}, offering a precise definition for an \emph{agent} is just as difficult as offering a precise definition for \emph{intelligence}. 
Wooldridge and Jennings also mention that this ambiguity may pose a danger for the term \emph{agent} to become a noise word in literature, used inconsistently.
Despite this ambiguity, however, our hope is to progressively refine our understanding of what agency means in the context of visual analytics.

\section{Conclusion}

We considered mixed-initiative visual analytics, where humans and computers collaborate to solve problems involving data.
Upon providing an overview of existing \acro{VA} frameworks, we proposed an agent-based framework to reason about the entire \acro{VA} pipeline.
We framed \acro{VA} settings as a set of human agents (e.g.\ analysts) and artificial agents (e.g. guidance engines) that interact with their environment (e.g. datasets and visualizations).
This framework draws parallels from the \acro{AI} literature, helping us to adapt a similar language and bridge some of the gaps between the two fields.
We offered three case studies from \acro{VA} litersture and discussed how this model can guide the design of future mixed-initiative \acro{VA} systems.

\section*{Acknowledgements}
The authors would like to thank Stylianos Loukas Vasileiou and Sunwoo Ha for their valuable feedback and conversations.
This material is based upon work supported by the U.S.\ National Science Foundation under grant numbers IIS-2142977, OAC-2118201, 1704018, and 2211845. 
%
%
%
%
%

%
%
%
%

\bibliographystyle{eg-alpha-doi}
\bibliography{main}



\end{document}